\documentclass[%
 jcp,
nofootinbib,
 amsmath,amssymb,
 table
preprint,%
]{revtex4-2}

\usepackage{graphicx}
\usepackage{dcolumn}
\usepackage{bm}

\usepackage[utf8]{inputenc}
\usepackage[T1]{fontenc}
\usepackage{mathptmx}
\usepackage{etoolbox}

\usepackage{hyperref}       
\usepackage{url}            
\usepackage{booktabs}       
\usepackage{amsfonts}       
\usepackage{nicefrac}       
\usepackage{microtype}      
\usepackage[table]{xcolor}         
\usepackage{amsthm}
\usepackage{bbm}
\usepackage{tablefootnote}

\usepackage{graphicx}
\usepackage{multirow}
\usepackage{siunitx}
\newcolumntype{Y}{S[table-format = 0.3{\tiny~$\pm0.00$}, table-align-text-post = true]}

\makeatletter
\def\@email#1#2{%
 \endgroup
 \patchcmd{\titleblock@produce}
  {\frontmatter@RRAPformat}
  {\frontmatter@RRAPformat{\produce@RRAP{#1\href{mailto:#2}{#2}}}\frontmatter@RRAPformat}
  {}{}
}%
\makeatother
\begin{document}

\preprint{arXiv:2609.18093}

\title[EESI]{Entropy Estimates from Stochastic Interpolants}
\author{Phillip M. Rauscher}
\email{p.rauscher@nyu.edu}
\affiliation{ 
Department of Chemical \& Biomolecular Engineering \\ Tandon School of Engineering, New York University}%

\date{\today}

\begin{abstract}
We present a general method for estimating entropy differences between arbitrary probability distributions using stochastic interpolants.
The marginal distributions which bridge the base and target obey a continuity equation, which allows a straightforward calculation of the entropy difference in terms of an inner product of the probability flow velocity and score fields.
This formulation has several advantages: (i) no computationally expensive divergence calculations of either field are required, (ii) the score field need not even be learned directly if model transferability is not required, and (iii) on-the-fly estimates are produced nearly for free during training.
When tractable base distributions are chosen (e.g. Gaussian chain, ideal gas, etc.), statistical thermodynamic entropies are then immediately recovered.
Notably, the analysis relies only on the definition of the Gibbs-Shannon entropy, rather than any particular statistical mechanical ensemble, so that generalized non-equilibrium entropies may be computed.
The method is demonstrated on several systems of increasing complexity: (i) a 40-dimensional Gaussian mixture model, (ii) the classical $XY$ model of $N$ spins arranged in one dimension, (iii) the 13-atom Lennard-Jones cluster, and (iv) an active Brownian polymer.
\end{abstract}

\maketitle

\section{Introduction}
\label{intro}

Generative models have shown great promise in the physical sciences for tasks such as material design \citep{Hoellmer2025,Zeng2025}, crystal structure prediction \citep{Zeng2026,Kilgour2026}, equilibrium conformational sampling \citep{Noe2019, Morehead2024,Albergo2024,Tan2025,Schebek2026,Havens2025,Gloy2025}, and free energy calculations \citep{Lu2023,Mate2024,Mate2025,He2025}.
These advances have further benefited from symmetry and geometry-aware formulations \citep{Kohler2020,Klein2023,Wu2025,Hoffmann2026} and large-scale pre-training efforts \citep{Klein2024,Tan2025}, so that the use of generative models in molecular science is becoming increasingly practical.

The methods mentioned above generally focus on sampling from equilibrium, or Boltzmann, distributions of the form:
\begin{equation}
    \rho(x) \propto \exp \left [-\frac{U(x)}{k_BT}\right ]
    \label{eq:boltz}
\end{equation}
where $U(x)$ is an energy function, $k_B$ is Boltzmann's constant, and $T$ the temperature.
The major challenges in this case are that the normalization constant for Eq.~\ref{eq:boltz} is not known \textit{a priori} and that the phase space of $x$ is high dimensional with (possibly) large energy barriers.
Even so, knowing the general form of $\rho(x)$ allows for a variety of mathematical manipulations and sophisticated computational strategies to be employed \citep{Tuckerman2023,Purohit2026,Marcello2026}.

But what if our system is not at equilibrium?
In such cases, the probability law in Eq.~\ref{eq:boltz} will not hold in general; indeed for strongly nonequilibrium systems, the temperature itself may not be well-defined.
The general dynamical structure of nonequilibrium systems relies not on a free energy, but on energy and entropy as separate quantities which generate the reversible and irreversible dynamics, respectively \citep{Grmela1997,Oettinger1997,Oettinger2005}.
Therefore, in order to apply generative models to arbitrary nonequilibrium systems, we require a means of estimating entropies directly without recourse to the Boltzmann distribution.
Here, we address this challenge within the framework of stochastic interpolants, enabling simulation- and divergence-free estimates of general Gibbs-Shannon entropies.

The main contributions are as follows:
\begin{enumerate}
    \item We formulate a novel expression for entropy changes along probability flows as an inner product of the velocity and denoising score fields, both of which may be learned without expensive divergence calculations. When one of the distributions is chosen appropriately, thermodynamic entropies follow immediately.
    \item We demonstrate data-driven, simulation-free methods for estimating entropy differences between arbitrary probability distributions and test transferability of learned models. We further show that these estimates are available for negligible cost \textit{during} training.
    \item We apply the method to several reference systems, including a classical spin model defined on a Riemannian manifold, $(S^1)^{\otimes N}$, with symmetry group $\text{Z}_2 \times \text{O}(2)$, which is efficiently trained using equivariant optimal transport. Other reference systems include Gaussian mixture models and a 13-atom Lennard-Jones cluster.
    \item We present the first estimate for the configurational entropy of a tangentially-driven active polymer, without introducing any auxiliary representations or approximations, and without applying any coarse-graining operations.
    \item We discuss the consequences of model overfitting or ``memorization'' \citep{Somepalli2023,Chen2024,Bonnaire2025} and demonstrate the convergence to exact results as dataset size increases.
\end{enumerate}

\section{Background}
\label{related}

The vast majority of methods for calculating thermodynamic potentials from simulations (apart from internal energy) are focused on free energies rather than entropies. 
This is because the Boltzmann distribution, Eq.~\ref{eq:boltz}, and its isobaric and/or grand canonical variations, allow for a wide array of mathematical manipulations.
Examples include umbrella sampling, metadynamics, adaptive biasing force methods, and basis function sampling \citep{Marcello2026,Purohit2026}.
On the other hand, thermodynamic entropies can be obtained by multicanonical methods, such as Wang-Landau sampling \citep{Wang2001,Wang2001b,Landau2004} or statistical temperature molecular dynamics \cite{Kim2006}, which are iterative methods designed to sample an artificial ``flat'' energy distribution.
These methods directly obtain the density of states, allowing for powerful temperature-transferable analysis, but struggle with convergence for large systems and nonequilibrium implementations have not been developed (to the best of our knowledge).

Taking a very different approach, \citet{Avinery2019} showed that entropies can be effectively estimated in a thermodynamically-agnostic manner with compression algorithms.
Although such an approach is clearly amenable to arbitrary data sets, the application to physical systems requires several intermediate steps, including choosing a particular representation, coarse-graining and discretizing the degrees of freedom, and generating ideally random and ordered systems for comparison.

Within the context of activate matter, a variety of methods have emerged to estimate entropy \textit{production}. 
These include the information-theoretic strategy of \citet{Ro2022}, who estimated the relative probabilities of forward and reverse trajectories by way of cross-parsing complexity and subsequently exploited fluctuation theorems to obtain local entropy production.
Such an approach, while elegant, still relies on results from stochastic thermodynamics, which all assume some form of thermalized system in the form of a constant-temperature heat bath, limiting their range of applicability.

A more general method applicable to steady-state conditions is that of \citet{Boﬀi2024}. 
Those authors analyzed a general stochastic equation of motion and studied the entropy change of the associated Fokker-Planck equation, developing a score-matching strategy enabling entropy production rate estimates.
The equations derived in that work are nearly identical to those we present below; the key difference is that our work analyzes an auxiliary dynamics within the context of a generative flow model (see Section \ref{sec:sis}), and therefore does not require that the system of interest obey any particular equation of motion, stochastic or otherwise.
On the other hand, entropy production is not immediately available within our framework, since we generally do not have access to a deterministic drift.

In the field of generative modeling, continuous normalizing flows \citep{Papamakarios2019,Albergo2022,Lipman2022} have shown excellent capability in molecular structure generation. 
As they also enable exact calculation of sample (log) likelihoods by the continuous change-of-variables formula, they also allow for robust importance sampling and reweighting to faithfully reproduce the Boltzmann distribution, Eq.~\ref{eq:boltz}; such models are aptly named ``Boltzmann generators'' \citep{Noe2019,Klein2024,Schebek2026}.
In fact, these methods already enable ensemble-free entropy calculations since the change in Gibbs-Shannon entropy of two distributions is nothing but the mean change in negative log-likelihood (NLL), although this connection is rarely made explicit.
One notable exception is the flow matching mutual information (FMMI) strategy of \citet{Butakov2025}, who also introduce a simulation-free method of evaluating the change-of-variables formula.
In both of the latter cases, one is required to evaluate the divergence of the probability flow velocity, which is computationally expensive.

\section{Method}
\label{method}

\subsection{Time-Dependent Gibbs-Shannon Entropy}

Consider an arbitrary probability density function parameterized by time, $\rho_t(x) : \mathbb{R}^d \rightarrow \mathbb{R}_{\geq 0}$, with $x \in \mathbb{R}^d$ and $t \in [0,1]$. 
The Gibbs-Shannon entropy associated with this distribution is defined:
\begin{equation}
    S_t = - \int_{\mathbb{R}^d} \rho_t(x) \ln \rho_t(x) dx
    \label{eq:gibbs-shannon}
\end{equation}
For a physical system of $N$ particles, the thermodynamic entropy is simply Eq. \ref{eq:gibbs-shannon} multiplied by the Boltzmann constant $k_B$ with $x = (q^{3N},p^{3N})$ the phase space coordinate.
So long as $\rho_t(x)$ varies sufficiently smoothly with $t$, we can calculate the difference in entropy between two distributions, say $\rho_1(x)$ and $\rho_0(x)$, by direct time integration:
\begin{equation}
    \Delta S = S_1 - S_0 = \int_0^1 \left (\partial_t S_t\right )dt
\end{equation}
Evaluating the time derivative of Eq.~\ref{eq:gibbs-shannon}, we have
\begin{equation}
    \partial_t S_t = - \int_{\mathbb{R}^d} \partial_t \rho_t (x) \left [ \ln \rho_t(x) + 1\right ]d x = - \int_{\mathbb{R}^d} \partial_t \rho \left ( \ln \rho \right ) d x
    \label{eq:entropy_deriv}
\end{equation}
where the second equality follows from the fact that total probability is normalized to unity at all times.
Note that we have made implicit the dependence on $x$ and $t$ where there is no risk of confusion.

\subsection{Stochastic Interpolants} \label{sec:sis}
As formulated by \citet{Albergo2023} a \textit{stochastic interpolant} is a stochastic process defined as follows:
\begin{equation}
    x_t = I(x_0,x_1,t) + \gamma(t)z
    \label{eq:si}
\end{equation}
where $I(0,x_0,x_1) = x_0$, $I(1,x_0,x_1) = x_1$, $\gamma(0) = \gamma(1) = 0$, $z \sim \mathcal{N}(0,I_d)$, and $(x_0,x_1) \sim \nu$, with $\nu$ a joint distribution that marginalizes on $\rho_0$ $(\rho_1)$ when integrated with respect to $dx_1$ $(dx_0)$. The random variable $x_t$ is governed by a probability density $\rho_t(x)$ that smoothly bridges $\rho_0(x)$ and $\rho_1(x)$, and obeys a continuity equation:
\begin{align}
    \partial_t \rho = - \nabla \cdot \left ( b \rho \right ) 
    \label{eq:transport}
\end{align}
where $b = b(t,x) = \mathbb{E} [ \dot{x}_t | x_t = x]$ is the probability flow velocity with the expectation taken over $\rho_t(x)$.
In practice such expectations are obtained in one of two ways.
First, one may sample $x_t$ directly by drawing samples of $(x_0,x_1) \sim \nu$ and $ z \sim\mathcal{N}(0,I_d)$, which is convenient for model training.
Second, if the velocity field is known (or parameterized), one may use a generative process $X_t$ (whose law coincides with that of $x_t$); this process solves the following ordinary differential equation:
\begin{align}
    \frac{d X_t}{dt} = b(t,X_t), \qquad X_{t=0} \sim \rho_0.
    \label{eq:ode}
\end{align}
It is also possible to realize the process $X_t$ as a solution to a stochastic differential equation with drift $b_F(t,x) = b(t,x) + \epsilon(t) s(t,x)$ where $\epsilon(t) > 0$ is a diffusion coefficient and $s(t,x)$ is the score:
\begin{equation}
    s(t,x) = \nabla \ln \rho = \gamma^{-1}(t) \times \mathbb{E} \left [ z | x_t = x\right ]
    \label{eq:score}
\end{equation}
In this case, Eq.~\ref{eq:transport} becomes a Fokker-Planck equation; the results which follow remain valid in either case.

\subsection{Entropy Differences}
We now substitute Eq.~\ref{eq:transport} into Eq.~\ref{eq:entropy_deriv}, and integrate by parts to find:
\begin{equation}
    \partial_t S_t = - \int_{\mathbb{R}^d} \rho \left ( b \cdot \nabla \ln \rho \right ) dx = - \mathbb{E} \left [ b \cdot s \right ]
    \label{eq:entropy_ibp}
\end{equation}
The total entropy difference is then found by integration over $t$, which can be freely exchanged with that over $x$ (by virtue of the regularity of $\rho_t(x)$, see \cite{Albergo2023}) to yield:
\begin{equation}
    \Delta S = - \int_0^1\mathbb{E} \left [ b \cdot s  \right ]dt= - \mathbb{E} \left [ \int_0^1 \left (b \cdot s \right ) dt \right ]
    \label{eq:entropy_si}
\end{equation}
This is the key result which enables us to efficiently estimate entropies (or more precisely, differences in entropy), since both $b(t,x)$ and $s(t,x)$ may be estimated, e.g. as neural networks, if we can draw samples from both $\rho_0$ and $\rho_1$.
Importantly, both fields may be learned without expensive divergence calculations, which greatly accelerates both training and sampling.
The application to physical systems is discussed later on.

The result given in Eq.~\ref{eq:entropy_ibp} is essentially identical to that of \citet{Boﬀi2024} (see their Eq. 19), but applied to the dynamics of a probability flow rather than a physical system; connections to that work also appear in the next section.

\subsection{Continuous Change of Variables Formula}

The result in Eq.~\ref{eq:entropy_ibp} is also closely related to the continuous change-of-variables formula used to obtain sample likelihoods in flow matching.
Using the identity $\nabla \ln \rho = \nabla \rho/\rho$ and then integrating by parts, we find:
\begin{equation}
    \partial_t S_t = \int_{\mathbb{R}^d} \rho \left ( \nabla \cdot b \right ) dx = \mathbb{E} \left [ \nabla \cdot b \right ]
    \label{eq:div_vel}
\end{equation}
\begin{equation}
    \Delta S = \int_0^1\mathbb{E} \left [ \nabla \cdot b  \right ]dt= \mathbb{E} \left [ \int_0^1 \left (\nabla \cdot b \right ) dt \right ]
    \label{eq:entropy_si_div}
\end{equation}
Eq.~\ref{eq:entropy_si_div} is a spatially-averaged version of the continuous change-of-variables formula, which gives the change in log-likelihood of a sample along a flow path \citep{Noe2019}.
The connection with entropy is therefore self-evident, but usually implicit.

While this formula is widely used, it is generally expensive to calculate and approximations (e.g. Hutchinson's trace estimator) suffer from slow convergence in high dimensions \citep{Kohler2020}.
Therefore, the inner product formulation of Eqs.~\ref{eq:entropy_ibp} and \ref{eq:entropy_si} offers significant computational savings.
However, it is important to note that the quantity $\int_0^1 (b \cdot s )dt$ cannot be used to obtain sample likelihoods: the right-hand sides of Eqs.~\ref{eq:entropy_ibp} and \ref{eq:div_vel} are only equal under the expectation, but do not have the same distribution.

In practice, the time integrals in Eqs.~\ref{eq:entropy_si} and \ref{eq:entropy_si_div} may be realized either by solving Eq.~\ref{eq:ode} (or the related SDE) for an ensemble of initial conditions, $X_0 \sim \rho_0$.
This enables the use of transferable models \citep{Mate2024,Klein2024,Tan2025}, which may be trained on one system, for instance at a particular density or system size, and used to generate samples from another related one.
However, if samples from the target distribution are available, Eqs.~\ref{eq:entropy_si} and \ref{eq:entropy_si_div} can also be evaluated in a more efficient generation-free manner by taking the expectation over $(x_0,x_1) \sim \nu, z \sim \mathcal{N}(0,I_d),$ and $t \sim U(0,1)$, as suggested by \citet{Butakov2025}.

\subsection{Imperfect Learning}

In general, the learned velocity and score fields, which we denote $\hat{b}(t,x)$ and $\hat{s}(t,x)$, respectively, include some approximation error, which is compounded by the inner product $\hat{b} \cdot \hat{s}$ in Eq.~\ref{eq:entropy_ibp}.
In general, we find that the score can be particularly challenging to learn.
In such cases, we can insulate our results from inaccuracies in $\hat{s}(t,x)$ by computing the entropy accumulation as follows:
\begin{align}
    \partial_t S_t = - \mathbb{E}[b \cdot s] &\approx -\mathbb{E}[\hat{b}(t,x) \cdot s(t,x)] = \frac{1}{\gamma(t)} \int_{\mathbb{R}^d} \hat{b}(t,x) \cdot \mathbb{E} \left [ z | x_t = x \right ] \rho_t(x)dx \nonumber \\
    & = \frac{1}{\gamma(t)} \mathbb{E}\left [ \hat{b}(t,x_t) \cdot{z}\right ]
    \label{eq:ent_dotz}
\end{align}
where the expectations are taken over $(x_0,x_1)\sim \nu$ and $z \sim \mathcal{N}(0,I_d)$ independently as usual.
Integration over time yields the entropy difference $\Delta S$ as a further expectation over $t \sim U(0,1)$.

The factor of $1/\gamma(t)$ in Eq.~\ref{eq:ent_dotz} leads to large variances near the endpoints.
To handle these numerically, we employ antithetic sampling as proposed by \citet{Albergo2023}.
In general, we find that Eq.~\ref{eq:ent_dotz} has higher variance than Eq.~\ref{eq:entropy_ibp} but significantly smaller bias, leading to more accurate results.
In fact, computational expense is markedly reduced since no score model needs to be trained at all.

The primary drawback of neglecting the score is a lack of transferability: the latent variable $z$ is only available when the law of the interpolant $\rho_t(x)$ is sampled via $x_t$, which depends explicitly on samples $x_1$; such samples may not be available for systems of interest.
On the other hand, the score model can be evaluated along the trajectories $X_t$ generated by Eq.~\ref{eq:ode} without reference to the latent variable.
One could imagine a hybrid approach in which the generated samples $X_1$ are used to evaluate Eq.~\ref{eq:ent_dotz}, but there are two difficulties with such a strategy. 
First, it exacerbates any inaccuracies in the velocity model, accumulating errors across entire trajectories; this is the reason why Boltzmann generators use a reweighting scheme following integration of Eq.~\ref{eq:ode} \cite{Noe2019,Schebek2026}.
Second, the interpolant $x_t$ depends on the joint distribution $\nu(x_0,x_1)$. 
Optimal transport or pre-alignment alters this distribution in manner that may not be the same for different systems, or even identical systems but with larger dimension (i.e. more particles). 
Therefore, inconsistencies may arise between the presumed velocity field and the true interpolant density $\rho_t(x)$.

\subsection{(Nearly) Free Estimates During Training}

Because the computational expense of calculating Eqs. \ref{eq:entropy_ibp} and  ~\ref{eq:ent_dotz} is so small, the predicted entropy change can be monitored during training.
Since forward passes of $\hat{b}(t,x)$ and $\hat{s}(t,x)$ and sampling of latent variable $z$ are already required for calculating the loss(es), any extra inner products then require only $O(d)$ operations.
If both velocity and score are trained, the two estimates can be compared to assess model consistency and in particular convergence of the score model.

The saturation of the entropy estimate might also be useful in determining when training should be stopped or modified. 
Intuitively, if the model is no longer able to add or remove information (i.e. change entropy), then the training parameters may no longer be appropriate.

\subsection{Physical Systems}

To connect with physical systems, we typically choose $x \in \mathbb{R}^{3N}$ as $N$ particle positions, $\rho_0(x)$ as a non-interacting reference system (e.g. ideal gas), and $\rho_1(x)$ taken from the system of interest.
Then upon evaluating Eqs.~\ref{eq:entropy_ibp} and \ref{eq:entropy_si_div}, we have $\Delta S = S_{\text{ex}}$, the excess entropy.
It is trivial to sample from ideal systems, while interacting ones can be accessed by standard simulations, e.g. Monte Carlo or molecular dynamics methods.
Note that the reference system need not be strictly ideal or de-correlated, but can instead be chosen for computational or theoretical convenience.
For example, to estimate the configurational entropy of an active macromolecules, we will take the reference system to be a polymer with bond stretching and bending interactions, but neglecting excluded volume; the partition function is more complicated, but still analytically tractable, enabling direct estimates of the microscopic entropy, while also making modle training more efficient.
Moreover, the phase space state $x$ need not be Euclidean or physical space, but can exist on general Riemannian manifolds \citep{Wu2025,Grenioux2025,Hoffmann2026} to describe spin models, which we will also demonstrate.

\subsection{Equilibrium and Nonequilibrium Ensembles}

Interestingly, the relationships derived above rely only on the definition of the Gibbs-Shannon entropy Eq.~\ref{eq:gibbs-shannon} and the standard properties of probability flows: no explicit invocation of any particular statistical mechanical ensemble is required.
As a result, the method may be used to compute \textit{nonequilibrium} entropies, which may prove valuable in the context of multiscale modeling since these quantities are the drivers of irreversible dynamics.
Indeed Eq.~\ref{eq:gibbs-shannon} has long been proposed as a general nonequilibrium entropy \citep{Jaynes1980,Oettinger2005}, including for situations where the standard assumptions of local equilibrium and/or thermalization are questionable.
For these reasons, we believe that this method, which we term ``entropy estimates from stochastic interpolants'' (EESI), may offer significant advantages over other methods which presuppose Boltzmann-distributed ensembles.

On the other hand, if we return to equilibrium ensembles, general free energies are easily calculated within this framework by standard thermodynamic relations \citep{Tuckerman2023}.
Concretely, the (Helmholtz) free energy difference between the two populations characterized by $\rho_0(x)$ and $\rho_1(x)$ follows from $\Delta F = \Delta U - T \Delta S$, where $\Delta U = U_1-U_0$ is the average energy change between base and target systems.
We will use these equilibrium free energies to validate the method before moving to nonequilibrium scenarios.

\section{Results}

We now apply the proposed method to a series of systems of increasing complexity, starting with three equilibrium cases where results can be validated against direct sampling, analytical formulae, or traditional thermodynamic integration.
We then apply the method on a well-studied active nonequilibrium system.
Architectural details and hyperparameters are reported in Appendix \ref{app:exp} and the results are summarized in Table \ref{tab:res}, with all quantities expressed in units of nats.
For simplicity, the same exact architecture and hyperparameters are used for learning both score and velocity fields except for the first example (Gaussian Mixture Model).
Emphasizing the efficiency of the method, all calculations were run on a desktop workstation (see Section \ref{sec:infra}).

\begin{table*}[b]
\centering
\caption{Comparative entropy difference between base (ideal) and target (real) systems ($\Delta S = S_1-S_0$) as calculated by the various methods described here. The reference values for the GMM, XY, and Lennard-Jones/Polymer systems are obtained by direct sampling, analytical solution, and thermodynamic integration, respectively. See Appendix \ref{app:exp} for details. Uncertainties in the means are reported as 95\% confidence intervals determined by bootstrap error analysis. *The generative inner product uses Eq.~\ref{eq:entropy_si} while the non-generative applies Eq.~\ref{eq:ent_dotz}.
}\label{tab:res}
\begin{ruledtabular}
\begin{tabular}{@{}llYYYY@{}}
\multirow{2}{*}{\textbf{Method}} &  & \textbf{GMM} & \textbf{XY Chain} & \textbf{Lennard-Jones} & \textbf{Active Polymer}\\ \cmidrule(lr){3-6}
 & & \small $16$ modes, $d=40$ & \small $d=10$  & \small $d=13\times3$ & \small $d=20\times3$ \\ \midrule
\textbf{Reference} & & -118.19 {\scriptsize $\pm 0.01$} & -5.144 & -34.17 {\scriptsize $\pm 0.02$} & -4.61 {\scriptsize $\pm 0.01$} \\
\midrule
\multirow{2}{*}{\textbf{Inner Product}}* & \small Generative & -114.7 {\scriptsize $\pm 0.4$} & -4.84 {\scriptsize $\pm 0.03$} & -33.9 {\scriptsize $\pm 0.2$} & -4.95 {\scriptsize $\pm 0.09$} \\ \cmidrule(lr){2-6} (EESI) & \small Non-Generative & -117.81 {\scriptsize $\pm 0.15$} & -5.12 {\scriptsize $\pm 0.07$}  & -34.14 {\scriptsize $\pm 0.11$} & -4.78 {\scriptsize $\pm 0.08$} \\
\midrule
\multirow{2}{*}{\textbf{Divergence}} & \small Generative & -117.70 {\scriptsize $\pm 0.03$} &  -5.19 {\scriptsize $\pm 0.03$} & -33.71 {\scriptsize $\pm 0.11$} & -4.81 {\scriptsize $\pm 0.07$}\\ \cmidrule(lr){2-6} & \small Non-Generative & -117.85 {\scriptsize $\pm 0.15$} & -5.33 {\scriptsize $\pm 0.04$} & -34.07 {\scriptsize $\pm 0.10$}  &  -4.66 {\scriptsize $\pm 0.06$} \\
\end{tabular}
\end{ruledtabular}
\end{table*}

\subsection{Gaussian Mixture Model}

We first consider the entropy of a Gaussian mixture model (GMM) with $n=16$ modes and dimension $d=40$; the values for the centers and covariances are exactly those given by \citet{Midgley2022} and also used by \citet{He2025}.
The base (or prior) distribution is the standard multivariate normal whose Shannon entropy is $S_0 = (d/2) \ln \left ( 2 \pi e \right)$.
Although no such analytical result is available for the GMM target density, the entropy can be accurately estimated by computing $S_1 = -\mathbb{E}_{x \sim \rho_1 }\left [\ln \rho_1(x) \right ]$ via direct sampling.
Details of model architecture and training are provided in Appendix \ref{app:gmm}.

The results are presented in Table~\ref{tab:res}, with most methods exhibiting errors of less than 0.5 nats, which falls within a commonly-used target accuracy for quantum chemical calculations (roughly 1 kcal/mol or 1.7 $k_BT$ at room temperature).
Such results are quite satisfactory given the rather simple network parameterizations and the lack of any reweighting or correction scheme. 
The only exception is the method that relies on the learned score (inner product, generative), which shows an error of about 3.5 nats (relative error of $\sim 3\%$); in this example, the score is particularly difficult to learn.
On the whole, the agreement with the exact results demonstrates that this method of computing entropy differences is valid in principle.

\subsection{Classical XY Model in One Dimension}

For a more physically-relevant example, we consider an open ``chain'' of $N$ interacting spins, i.e. the one-dimensional XY model \citep{Mattis1985}.
This choice is convenient as the canonical partition function is exactly solvable, providing a rigorous metric for method accuracy. 
Furthermore, system configurations can be obtained without simulations by sampling the von Mises distribution.
The base distribution is a series of non-interacting spins distributed uniformly on the unit circle.

The system also provides a test bed for model transferability.
Concretely, learning the velocity/score fields as finite-range graph neural networks (GNNs), it is possible to apply models trained at one value of $N$ to systems with different sizes.
The system is also a simple case of a Riemannian manifold: each spin is a unit vector in 2D, so the phase space domain is the topological space $(S^1)^{\otimes N}$, with a distribution function that is invariant under the symmetry group $\text{Z}_2 \times \text{O}(2)$.
Technical details of the system and associated models are given in Appendix \ref{app:xy}.

\begin{figure}[b]
    \centering
    \includegraphics{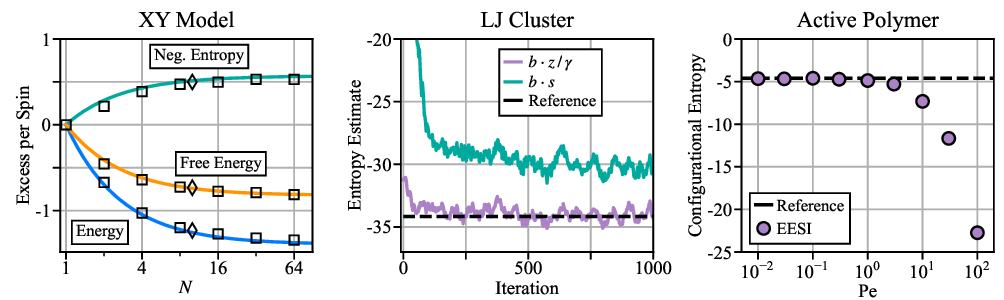}
    \caption{Experimental results. Left: Excess energy (blue), negative entropy (teal), and free energy (orange) per spin for the classical XY model with varying $N$. The models were trained from data with $N=10$ (diamonds) and then used to estimate properties for other values (squares). Middle: On-the-fly entropy estimates (smoothed) for the 13-particle Lennard-Jones cluster, showing the convergence of Eq.~\ref{eq:entropy_ibp} (teal) and Eq.~\ref{eq:ent_dotz} (violet) to the thermodynamic integration reference value (dashed black) during fine-tuning of a pre-trained model \citep{Klein2023}. Right: Estimated configurational entropy for a tangentially active polymer (violet) with varying activity (P\'{e}clet number). The dashed black line is the $\text{Pe}=0$ reference obtained by thermodynamic integration with respect to the ideal chain (i.e. without excluded volume).}
    \label{fig:results}
\end{figure}

A system with $N=10$ spins is chosen for model training and the resulting entropy differences are again fairly accurate, with errors of at most 0.3 nats for all flavors of entropy estimation (see Table \ref{tab:res}).
The methods are also show good transferability in predicting both energy and entropy as illustrated in Fig.~\ref{fig:results} (left), with no loss of accuracy upon increasing to $N=64$.
In contrast, the results for smaller systems, especially $N=2$, are less impressive. 
This is likely a boundary effect: we chose not to encode any notion of lattice position in our networks, so the unique nature of the chain ends must be learned independently.
The small system of $N=2$ is simply two chain ends, so the larger errors are not unexpected in this case.
A more expressive model (or one with greater inductive bias) would likely improve these results for small systems.

\subsection{13-Particle Lennard-Jones Cluster}

We now turn to a standard test system for generative models of molecular systems: a cluster of 13 Lennard-Jones particles at thermal equilibrium.
The system includes a small harmonic restraining potential to keep the cluster intact; in the absence of the Lennard-Jones potential, this results in a collection of independent harmonically restrained particles (with center-of-mass removed).

We use the same training data and methods as \citet{Klein2023} and learn the velocity field by fine-tuning their model, which is an E(n)-equivariant graph neural network (EGNN, \cite{Satorras2021}); the model for the score field has the same structure and is initialized with small random weights.
For this system, the gains in computational efficiency become especially apparent, as the divergence calculation is approximately 50 times slower than the inner products.
To assess the accuracy of EESI, we conduct a thermodynamic integration to obtain the free energy difference $\Delta F$ between the reference and interacting systems.
The entropy difference then follows as $\Delta S = -(\Delta F - \Delta U)/T$ where $\Delta U$ is the change in potential energy between the states and $T$ is the temperature. 
For a more detailed description of the methods, see Appendix \ref{app:lj}.

The results given in Table~\ref{tab:res} again show rather reasonable agreement with errors below 0.5 nats in all cases, again without any reweighting or importance sampling scheme.
In the middle panel of Figure~\ref{fig:results}, we show on-the-fly estimates obtained during model fine-tuning, with accurate values obtained after as few as 500 batch iterations through Eq.~\ref{eq:ent_dotz}.
The convergence of the estimate from Eq.~\ref{eq:entropy_si} is slower as the score field is being trained from scratch, although the final results are still rather accurate.
These findings suggest that pre-trained transferrable flow models \citep{Klein2024,Tan2025} can be easily adapted to provide entropy estimates.

\subsection{Tangentially Active Brownian Polymers}

The final case we consider is a true nonequilibrium system which does not obey a Boltzmann distribution (or any other known analytical form) and is a well-studied model for active biological macromolecules \citep{Bianco2018}.
The system is a linear polar polymer with $N=20$ beads undergoing Brownian dynamics (neglecting hydrodynamics).  Bonded and excluded volume interactions are described by conservative forces, while activity is introduced in the form of tangential driving forces.
Each bead (except those at the chain ends) is subject to an ``active'' force of magnitude $f_{a}$ directed along the vector joining its two neighboring beads.
The relative strength of the activity is characterized by the P\'{e}clet number $\text{Pe}=f_a b/k_BT$ where $b$ is the equilibrium bond length and $k_BT$ is the thermal energy scale.
The system is notable for its activity-induced collapse into globule-like states at high $\text{Pe}$.

We generate polymer conformations by Brownian dynamics simulations, removing translational degrees of freedom by shifting the ``tail'' bead to the origin.
The base distribution is a polymer without any excluded volume interactions or activity, but still incorporating bonded potentials.
In addition, we add an angle bending potential to more closely match the chain dimensions of the prior/base and target distributions, which greatly reduces transport cost/distance.
Although this prior is more complex than a simple ideal Gaussian or freely-jointed chain, the partition function and configurational entropy are still available analytically and the reduced transport distance allows for faster training and improved accuracy.
This choice of prior also allows for an equilibrium ($\text{Pe} = 0$) benchmark via standard thermodynamic integration, which is reported in Table \ref{tab:res}.
The methods are described in detail in Appendix \ref{app:tap}.

We again use an EGNN model to parameterize the velocity/score fields and provide additional inductive bias by including the bead position (or index) along the chain as an additional node feature and introducing a Boolean-type flag indicating a covalent bond as an extra edge feature.
To reduce training demands, we use the validated initial $\text{Pe}=0$ models as warm-starts for other values of $\text{Pe}$, ranging from $10^{-2}$ to $10^2$.

At low $\text{Pe}$, the configurational entropy of the polymer is largely unaffected by the activity, adopting equilibrium-like conformations.
However, when $\text{Pe}>1$, large differences emerge, with entropy being strongly reduced.
Note that this is not the entropy \textit{production} rate, which has been extensively discussed in the literature, but is instead the entropy \textit{difference} between the active and passive ensembles, quantifying the thermodynamic restoring force.
At this point, the scaling behavior of the configurational entropy change as a function of $\text{Pe}$ cannot be clearly identified; we leave this topic to a future investigation.

\begin{figure}[htb]
    \centering
    \includegraphics{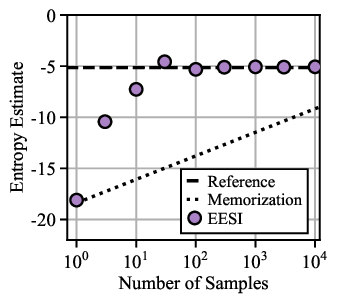}
    \caption{Entropy estimates obtained for the XY model with varying training data size. The dashed black line is the analytical result while the dotted line is the expected result in the case of memorization. All results are relative to the ideal (uniform) distribution and are expressed in nats.}
    \label{fig:ablation}
\end{figure}

\subsection{Learning vs. Memorization}

In data-limited regimes, diffusion and flow models are known to ``memorize,'' i.e. simply reproduce their training data during sampling, rather than generate truly novel samples \citep{Somepalli2023,Chen2024,Bonnaire2025}.
Since we anticipate operating in such a regime in chemical physics applications, it is worthwhile to comment on this behavior as it pertains to entropy estimation.
In particular, we return to the XY model introduced above and retrain our networks under varying sample sizes, with the results shown in Fig~\ref{fig:ablation}.

If the training data contains $M$ samples, then we expect that a model which has memorized this data yields an entropy estimate of the Boltzmann type, $S = -k_B \ln M$, which is plotted alongside the numerical data and the analytical entropy difference.
For $M=1$, the model is clearly memorizing the single sampled state.
This memorization quickly fades, so that the ``true'' entropy is recovered with as few as 100 samples.
Of course, the nature of this decay is dependent on the system and the model under consideration, but it is encouraging that a minimally expressive model and a rather limit dataset can still avoid memorization and produce accurate entropy estimates.

\section{Discussion and Conclusions}

We have presented a method for estimating entropy differences between physical systems, termed EESI.
The method is entirely data-driven and does not rely on any \textit{a priori} knowledge concerning the form of the distribution function, allowing for the study of nonequilibrium systems, such as active, driven, or flowing molecular ensembles.

There are two primary limitations of the EESI method.
First, the score field can be rather difficult to learn in practice as its objective has a large variance.
Pervious work addressed improved the training signal using target score matching \citep{DeBortoli2024,He2025}, which uses known values of the score at the end points.
At equilibrium, the score at the two end states is easily obtained as the gradient of the potential energy, i.e. the forces. 
However, no such relationship holds out-of-equilibrium for the target distribution as it is not Boltzmann-distributed.

It may, however, be possible to improve the score training by careful parameterization of the stochastic interpolant such that the score and velocity are directly related to one another \citep{Albergo2023}.
It may also be possible to use the equivalence of Eqs.~\ref{eq:entropy_ibp} and~\ref{eq:div_vel} as proposed by \citet{Boﬀi2024}.
Since $\nabla\cdot b$ tends to have much lower variance than $b\cdot s$, it can serve as an additional training target for the score.
This could be further integrated in an active learning framework, in which the time intervals of greatest uncertainty in $b\cdot s$ are refined by a small number of divergence calculations.
Of course, as mentioned above, learning the score is only truly necessary if the flow model is intended to be transferable. 

A somewhat related challenge is the inability to re-weight generated samples in nonequilibrium conditions.
This reweighting (importance sampling) is a key step in obtaining accurate equilibrium ensembles and free energies \citep{Noe2019,He2025}, but again relies on a known form of the probability density, even if un-normalized, as this provides a tractable log-likelihood.
Resolving this issue is a formidable challenge and is likely to require system-specific strategies.

\begin{acknowledgments}
Financial support for this work was provided by the Tandon School of Engineering at New York University. I am grateful to Professor Mark E. Tuckerman and Dr. Adam Lahouari for helpful discussions and feedback on this manuscript.
\end{acknowledgments}

\section*{Data Availability Statement}

The data that support the findings of this study are be openly available via the Open Science Framework, reference number "wf54q" at \href{https://osf.io/wf54q}{osf.io/wf54q}.
The PyTorch implementation of EESI is available at \href{https://github.com/Rauscher-Group/eesi.git}{github.com/Rauscher-Group/eesi}.
The codes for the Lennard-Jones and active polymer systems are freely available at \href{https://github.com/Rauscher-Group/LJClusterTI.jl.git}{github.com/Rauscher-Group/LJClusterTI.jl} and \href{https://github.com/Rauscher-Group/ActivePolyBD.jl.git}{github.com/Rauscher-Group/ActivePolyBD.jl}, respectively.

\appendix

\section{Systems, Model Architectures, and Training Details}\label{app:exp}

\subsection{Gaussian Mixture Model (GMM)}\label{app:gmm}

We first estimate the entropy of a Gaussian mixture in 40-dimensional space. 
Our implementation is exactly that of \citet{Midgley2022} and \citet{He2025}: 16 components, means $\sim U(-2, 2)$, std $=$ \texttt{softplus}$(-3)$, random seed 10.
The velocity fields are parameterized by a multilayer perceptron (MLP) with 4 hidden layers of width $d_h = 512$ with SiLU activation functions \citep{Elfwing2017}.
The score field has two hidden layers with $d_h = 3072$; for this system we found that wider (rather than deeper) networks performed best for parameterizing the score.
The input features are the $d=40$ dimensional position vector concatenated with a learned time embedding of size $d_h$.
The time embedding is output from an additional MLP with two hidden layers of size $d_h$ and a SiLU activation.
The input is a standard positional encoding \citep{Vaswani2017} of the time, $t$, also of size $d_h$.

We train the model using the Adam optimizer \citep{Kingma2014} with a batch size of 1000 for 100,000 iterations with a learning rate of $10^{-4}$ followed by a fine-tuning of 20,000 iterations with a learning rate of $10^{-5}$.
For each batch, fresh samples are drawn from both the base and target distributions.
This is in contrast to later experiments which use a static set of target samples.

\subsection{Classical XY Model in One Dimension}\label{app:xy}

The system is a ``chain'' of $N$ spins placed on a regular one-dimensional lattice \citep{Mattis1985}.
Each spin $\hat{\textbf{s}}_i$ is a 2D unit vector, which may be represented as a point on the unit circle $S^1$.
The spins interact with nearest neighbors via the dimensionless potential:
\begin{align}
    \beta H = -J\sum_{i=1}^{N-1} \hat{\textbf{s}}_i \cdot \hat{\textbf{s}}_{i+1} = -J\sum_{i=1}^{N-1} \cos(\theta_{i+1} - \theta_{i})
    \label{eq:xy_potential}
\end{align}
with $\theta_i$ the angle between an arbitrary fixed axis and the $i^{\text{th}}$ spin.
We use units such that $\beta = 1$ with coupling parameter $J=2$, which plays the role of an inverse effective temperature.
The canonical partition function factorizes immediately in terms of the angle differences, $\phi_i = \theta_{i+1} - \theta_{i}$, leading to the exact results:
\begin{align}
    Z &= \int_{-\pi}^{\pi}  e^{J\sum_{i=1}^{N-1} \cos(\theta_{i+1} - \theta_{i} )} d\theta^N = 2 \pi \left [ \int_{-\pi}^{\pi} e^{J\cos\phi } d \phi \right ]^{N-1} \nonumber  \\ 
    &= (2\pi)^N (I_0(J))^{N-1} = Z_{\text{id}} \times (I_0(J))^{N-1} \label{eq:XY_partition}
\end{align}
\begin{align}
    \Delta U/N = \frac{J}{N} \frac{\partial \ln (Z/Z_{\text{id}})}{\partial J} = -J \left (1 - \frac{1}{N} \right ) \frac{I_1(J)}{I_0(J)}
\end{align}
\begin{align}
    \Delta S/N = (\Delta U-\Delta F)/N =  \left (1 - \frac{1}{N} \right ) \left [ \ln I_0(J)- J\frac{I_1(J)}{I_0(J)} \right ] 
\end{align}
where $\Delta U = U_1 - U_0$ is the average energy difference between the interacting ($J=2$) system and the non-interacting ($J=0$) system, $\Delta F = -(1/\beta)\ln (Z/Z_{\text{id}})$ is the free energy difference of the same, and $I_{\nu}(x)$ is the modified Bessel function of the first kind of order $\nu$.

Notice that the probability density (i.e. Boltzmann factor) found in square brackets in the third term of Eq.~\ref{eq:XY_partition} is proportional to the von Mises distribution. 
Therefore, we generate system configurations by choosing a random orientation for the first spin, $\theta_1 \sim U(-\pi,\pi)$ and then $N-1$ samples from the von Mises distribution for angle differences, $\phi_i \sim \text{vonMises}(J)$, followed by cumulative summation.

The domain of the XY model may be considered as $N$-fold product space $(S^1)^{\otimes N}$ which forms a Riemannian manifold, although the metric tensor is unity which simplifies the situation considerably.
The stochastic interpolant for such manifolds is written in terms of the exponential and logarithmic maps as described by \citet{Wu2025} and further utilized by \citet{Grenioux2025,Hoffmann2026} for handling periodic particle systems.
\begin{equation}
    x_t = \text{Exp}_{x_0} \left (t \,\text{Log}_{x_0}x_1 + \gamma(t) z \right )
\end{equation}
where we take $ \gamma(t) = 0.2\sqrt{t(1-t)}$. 
Including the noise inside the exponential map ensures that the interpolant remains on the manifold at all times, even when perturbed from the geodesic joining $x_0$ and $x_1$.
A similar formulation was used implicitly by \citet{Mate2024} and \citet{Mate2025}.
Accordingly, the latent variable $z \sim \mathcal{N}(0,I_{d})$ is defined in the tangent space, $z \in T_{x_0}M$.
For the simple geometry here, this tangent space is just $(\mathbb{R}^1)^{\otimes N}$.

\subsubsection{Equivariant Minibatch Optimal Transport}

To improve training and sampling efficiency, we align base/target samples by minimizing transport cost within the symmetry group as described by \citet{Klein2023}.
The interaction potential, Eq.~\ref{eq:xy_potential}, is invariant under rotations and reflections in two dimensions, i.e. $O(2)$ symmetry.
It is not fully permutation invariant owing to the fixed nearest-neighbor interactions and the open chain ends, however it does possess a head-to-tail symmetry, i.e. it is invariant on reversing the order of the summation, a $Z_2$ symmetry.

With these considerations, it is easy to search the complete symmetry group to find the optimal alignment between any two samples.
We apply chain reversal and/or spin reflection to the base sample (leading to four realizations) and then find the optimal rotation that minimizes the transport distance to the target, measured as a Euclidean distance (see below) for each of the realizations, choosing the best configuration among the four.

To optimally align samples, we minimize the chordal (or Euclidean) distance between pairs of spins in the base and target samples.
Concretely, we define the angle difference as $\psi_i = \theta_i^{\text{target}} - \theta_i^{\text{base}}$, then search for a rotation of the base sample spins, $\delta$, that minimizes the following objective:
\begin{align}
    C = \sum_{i=1}^{N} \left [ 1-\cos ( \psi_i - \delta )\right ]
\end{align}
By direct differentiation and some trigonometric identities, it is straightforward to show than the optimal value of $\delta$ is the circular mean, $\overline{\psi}$:
\begin{align}
    \underset{\delta}{\arg\min} \,\, C = \overline{\psi} = \text{atan2}\left ( \sum_i \sin (\psi_i) \, , \, \sum_i \cos (\psi_i) \right ) 
\end{align}
where the second equality may be understood by direct differentiation and the use of trigonometric identities.

We intend to train the velocity and score fields at one particular value of $N$ and then apply them to estimate entropies for other systems. 
Therefore, we do not apply minibatch optimal transport methods as this would alter the joint target/base distribution $\nu(x_0,x_1)$ in a manner that depends on system and batch size, as discussed in the main text.

\subsubsection{1D Static Graph Neural Network}

The velocity and score fields are parameterized by an equivariant graph neural network (EGNN) inspired by that of \citet{Satorras2021}, but simplified and streamlined for the low dimensional system here and taking account of its slightly different symmetries.
First, we featurize the interpolant time by a Fourier expansion with logarithmically spaced frequencies in the range $1 \leq \omega \leq 30$ and then pass it through a learned embedding:
\begin{equation}
    \textbf{e}_t = f_t \left (\left [ t \oplus \left (\cos \omega_k t \right )\oplus \left (\sin \omega_k t \right )\right ] \right )
\end{equation}
with $\oplus$ indicating concatenation. 
Each equivariant graph convolution (EGC) layer of the model performs the following sequence of operations:
\begin{align}
    \phi_{ij}^l &= \texttt{wrap} \left ( \theta_{j}^l - \theta_i^l \right ) \\
    \textbf{e}_{ij}^l &= \left [ |j-i|^{-1}, \cos(\phi_{ij}^l), \cos(2\phi_{ij}^l), ...,\cos(n_e\phi_{ij}^l)\right ] \\
    \theta_i^{l+1} &= \theta_i^l + \sum_{j \in \mathcal{N}(i)} \phi_{ij}^l f_e \left (\textbf{e}_{ij}^l \oplus \textbf{e}_t \right )
\end{align}
where $\texttt{wrap}(x) = x - (2\pi)  \times\lfloor x/2\pi \rfloor - \pi$ and the sum in the angle update runs over the neighbors of each spin, which are taken as those spins separated by at most two lattice spacings.
The order of the expansions for $\textbf{e}_t$ and $\textbf{e}_{ij}^l$ is taken as eight.

The functions $f_{t/e}$ are parameterized by MLPs with SiLU activations. $f_t$ has one hidden layer with hidden and output dimension of 32 while $f_e$ has three hidden layers with hidden dimension 16 and scalar output.
The full EGNN model uses three EGC layers and returns an output for each lattice site as:
\begin{equation}
    b_i(t) = \theta_i^3 - \theta_i^0
\end{equation}
Since the features $\phi_{ij}^l$ and $|j-i|^{-1}$ are equivariant under rotation, reflection, and chain reversal, the successive angle updates are also equivariant, as is the vector field output.
Note also that the angle updates and resulting velocity are \textit{not} wrapped into the canonical region since they exist on the tangent space of the manifold, which is $\mathbb{R}^N$ as mentioned previously.
Each EGNN model has a total of 3488 parameters.

\subsubsection{Training and Hyperparameters}

Samples from the base distribution were drawn on-demand during training iterations.
To mimic a data-constrained scenario, the target distribution samples ($N=10$) were limited to a total of 10,000 configurations which were generated prior to the training.
The velocity and score models were trained with a batch size of 256 for 20,000 iterations via the Adam optimizer with a learning rate of $10^{-4}$ followed by 5,000 iterations at a rate of $10^{-5}$.

To sample from the model, we integrate the ODE \ref{eq:ode} using the Heun midpoint method with a discretization of 100 steps.
Entropy estimates are computed as the average of $100 \times 256 = 25,600$ samples with 95\% confidence intervals determined by bootstrap error analysis.

\subsection{Lennard-Jones Cluster}\label{app:lj}

We consider the alchemical free energy of a small cluster of $N=13$ Lennard-Jones particles.
The dimensionless system Hamiltonian is written:
\begin{align}
    \beta H = \frac{1}{2} \sum_{\langle i,j\rangle} \left [ \left ( \frac{r_m}{r_{ij}} \right )^{12} - 2 \left ( \frac{r_m}{r_{ij}} \right )^6 \right ] + \beta H_0
    \label{eq:lj}
\end{align}
where $r_m$ is the minimum energy distance (taken as the unit of length), $r_{ij} = |\bm{r}_i - \bm{r}_j|$ is the distance between particles $i$ and $j$, $\beta H_0$ is a reference Hamiltonian, and the sum is taken over all pairs of particles such that $i < j$.
The latter is a simple harmonic restraint that keeps the particles close to the cluster center-of-mass:
\begin{align}
    \beta H_0 = \frac{1}{2} \sum_{i=1}^{N} \left \| \bm{r}_i - \frac{1}{N}\sum_{j=1}^{N} \bm{r}_j\right \|^2
    \label{eq:harmonic}
\end{align}
We closely follow the work of \citet{Klein2023} for model design and training, using the same datasets and checkpoints provided by those authors \citep{Klein2025_OSF}, though sampling only $10^6$ configurations rather than the entire ensemble.
We implement the same model architecture using the reported weights as a ``warm start'' for training the velocity network. 
The score network has the same structure and size but has weights initialized to small random values.
The base distribution is a centered Gaussian distribution, which is exactly the Boltzmann density associated with $\beta H_0$, allowing for a direct comparison with thermodynamic integration results (see below).

The equivariant optimal transport method of \citet{Klein2023} was used with a batch size of 256.
Training was carried out using the Adam optimizer with a learning rate of $10^{-3}$ for 10,000 iterations followed by 10,000 iterations at a rate of $10^{-4}$.
Model parameter fluctuations were suppressed by an exponential moving average (EMA) scheme, with a damping parameter of 0.999.

\subsubsection{Monte Carlo Simulations and Thermodynamic Integration}

To obtain a ``ground truth'' entropy for comparison, we conduct additional simulations for thermodynamic integration \citep{Tuckerman2023}.
We use a nonlinear scaling of the Hamiltonian as proposed by \citet{Shirts2005} and \citet{Steinbrecher2007} which includes an additional control parameter, $\lambda$, such that $H(\lambda=0) = H_0$ and $H(\lambda=1) = H$ as given in Eq.~\ref{eq:lj}:
\begin{align}
    s_{ij}^{(\lambda)} &= \alpha(1-\lambda) + \left ( r_{ij}/r_m\right )^{6} \\
    \beta H_{\lambda} &= \beta H_0 + \frac{\lambda}{2} \sum_{\langle i,j\rangle} \left [ \left ( s_{ij}^{(\lambda)} \right )^{-2} - 2 \left ( s_{ij}^{(\lambda)} \right )^{-1} \right ]
\end{align}
with $\alpha = 1/4$.
The free energy then follows as $\Delta F = \int_0^1 \langle \partial_{\lambda} H_{\lambda} \rangle_{\lambda} d\lambda$, where the angled brackets indicate an ensemble average under the Boltzmann distribution associated with $H_{\lambda}$.
We divide the interval $\lambda \in [0,1]$ into 40 evenly-spaced intervals and evaluate the integral numerically via Simpson's rule.

We obtain samples from the various $H_{\lambda}$ by Metropolis Monte Carlo.
Trial moves were implemented as particle displacements chosen uniformly in $[-d,+d\,]^3$ with the range $d$ dynamically adjusted during the first $10^5$ cycles of equilibration (one cycle = one attempt per particle) to achieve a target acceptance rate of $\sim$50\%.
This was followed by $5\times 10^5$ further equilibration cycles and $10^6$ production cycles recording values of $H_{\lambda}$ and $\partial_{\lambda} H_{\lambda}$ every 100 cycles.
The procedure above was repeated 32 times with different random number generator seeds for each $H_{\lambda}$ to reduce statistical uncertainty.

\subsection{Tangentially Active Polymer}\label{app:tap}

To apply EESI to a true nonequilibrium (non-Boltzmann) system, we consider a polymer undergoing Brownian dynamics with tangential activity, adopting a simplified version of the model of \citet{Bianco2018}.
The non-dimensional stochastic equation of motion for the $i^{\text{th}}$ particle along the chain is written:
\begin{equation}
    d\textbf{r}_i = - \nabla_{\textbf{r}_i} U(\{ \textbf{r}\}) \,dt + \sqrt{2} \,\, d\bm{W} + \text{Pe} \cdot \hat{\mathbf{s}}_i \,\,dt
    \label{eq:sde_tap}
\end{equation}
where $U$ is the conservative potential energy of the system, $ d\bm{W}$ is an increment of the Wiener process, $\text{Pe}$ is the P\'{e}clet number characterizing the strength of the active force, and $\hat{\textbf{s}}_i = \left ( \textbf{r}_{i+1}-\textbf{r}_{i-1}\right )/ \left | \textbf{r}_{i+1}-\textbf{r}_{i-1}\right |$ is the central difference tangent (unit) vector centered on the $i^{\text{th}}$ particle.
As this tangent vector is not defined for the chain ends, those particles are left passive.

The conservative potential is the sum of bonded and excluded volume contributions:
\begin{align}
    U &= U_{b} + U_{nb} \\
    U_b &= \frac{k}{2} \sum_{i=1}^{N-1} \left ( \left | \textbf{r}_{i+1} - \textbf{r}_{i} \right | - 1 \right )^2 \\
    U_{nb} &= \frac{k}{2} \sum_{\langle i,j\rangle} \mathbbm{1}_{|i-j|>1} \times\left ( r_{ij} - 1 \right )^2 \times \Theta(1-r_{ij})
\end{align}
where $k=100$ and $r_{ij} = |\textbf{r}_i - \textbf{r}_j|$ is the distance between particles $i$ and $j$.
Within the nonbonded potential, the indicator function ensures that only nonbonded particles are included in the sum over pairs, while the Heaviside function $\Theta(x)$ enforces a short-range cutoff at $r_{ij}=1$, leading to a purely repulsive (excluded volume) interaction.

\subsubsection{Brownian Dynamics Simulations and Thermodynamic Integration}

The equation of motion Eq.~\ref{eq:sde_tap} is integrated by the Euler-Maruyama method with $dt = 0.001$.
We initialize the polymer chains as self-avoiding walks starting at the origin and simulate for $1.1\times10^8$ timesteps, discarding the first $10^7$ as equilibration time.
Ten replicas are run at each value of Pe to reduce statistical uncertainty and ensure uncorrelated samples.

For Pe$=0$, the polymer is entirely passive and therefore obeys the standard Boltzmann distribution.
To estimate the free energy of this system, we again conduct a thermodynamic integration.
Since the excluded volume potential is non-singular at the origin, we adopt a direct coupling:
\begin{equation}
    H_{\lambda} = U_b + \lambda U_{nb}
\end{equation}
which allows us to calculate the free energy difference associated with the introduction of excluded volume interactions.
Note that the partition function of the ``bare'' Hamiltonian is available analytically:
\begin{align}
    Z_b &= \int d\text{r}^N e^{- U_b} = \int d\text{r}^N \exp \left [ -\frac{k}{2} \sum_{i=1}^{N-1} \left ( \left | \textbf{r}_{i+1} - \textbf{r}_{i} \right | - 1 \right )^2 \right ] \nonumber \\
    &= V \left \{ \int d\bm{Q} \exp \left [ -\frac{k}{2} \left ( |Q| -1\right )^2 \right ] \right \}^{N-1} = V \left [ 4 \pi \int_0^{\infty} Q^2 e^{-\frac{k}{2} (Q-1)^2} dQ\right ]^{N-1} \nonumber \\ 
    &= V \left [ \frac{(1+k)(1+\text{erf}(\sqrt{k/2})}{(k/2\pi)^{3/2}} + \frac{2e^{-k/2}}{(k/2\pi)} \right ]^{N-1}
    \label{eq:partition_bond}
\end{align}
where $\text{erf}(x)$ is the error function and in the second line we have factorized the Boltzmann weight into independent terms by introducing the connector vectors $\bm{Q} = \textbf{r}_{i+1} - \textbf{r}_{i}$ and integrating the chain end position $\textbf{r}_1$ over all space.
The modified Hamiltonian is sampled by the same Brownian dynamics simulations described above, logging the values of $H_{\lambda}$ and $\partial_{\lambda} H_{\lambda}$ every 10,000 time steps, with ten independent simulations conducted for each $H_{\lambda}$.
A total of 25 values of $\lambda$ were considered ranging from 0 to 1 with values distributed uniformly in $\ln(1+\lambda)$.

\subsubsection{Base Distribution and Optimal Transport}

The equation of motion Eq.~\ref{eq:sde_tap} is equivariant under all rotations and reflections, but is \textit{not} symmetric under chain reversal as in the $XY$ model studied earlier.
Therefore, to minimize transport distance within the appropriate symmetry group E(3), we first place the ``tail'' chain end at the origin, apply a point reflection to generate two realizations, apply the Kabsch algorithm to each, and then retain the one with minimal squared Euclidean distance to the target sample. 
The associated cost matrix is then used to match target and sample configurations within each batch.

Although the system defined in Eq.~\ref{eq:partition_bond} is easily related to the passive system, it leads to much more compact polymers and as a result, large transport distances.
To mitigate this, the base distribution is modified to include an angle bending potential so that the polymer dimensions, i.e. end-to-end distance, remain similar in the base and target distributions:
\begin{equation}
    U_{\text{ang}} = \frac{\gamma}{2} \sum_{i=2}^{N-1} \left ( \cos\theta_{i} -\kappa\right )^2
\end{equation}
where $\theta_{i}$ is the angle between bond vectors $\textbf{r}_{i,i-1}$ and $\textbf{r}_{i+1,i}$, with $\kappa$ the cosine of the equilibrium angle and $\gamma$ a spring constant.
The partition function may still be obtained analytically for this model, 
\begin{align}
Z_{\text{base}} &= \int d\text{r}^N e^{- U_b-U_{\text{ang}}} = V \left [ 4 \pi \int_0^{\infty} Q^2 e^{-\frac{k}{2} (Q-1)^2}dQ \right ]^{N-1} \left [ \frac{1}{2} \int_{-1}^{1} e^{-\frac{\gamma}{2} (u - \kappa)^2} du \right]^{N-2} \nonumber \\
    &= Z_b \times Z_{\text{ang}}
\end{align}
where the second equality is obtained by take $\bm{Q}_i$ as the polar axis for the solid angle integration of $\bm{Q}_i$ along with a change of variables $u = \cos \theta_i$, noting that the first bond vector integration is unaffected by the bending potential.
Here the angular partition function is:
\begin{equation}
    Z_{\text{ang}} = \left \{ \frac{1}{2} \sqrt{\frac{\pi}{2\gamma}} \left [ \text{erf}\left ( \sqrt{\frac{\gamma}{2}} ( \kappa+1) \right )- \text{erf} \left ( \sqrt{\frac{\gamma}{2}} ( \kappa-1) \right  ) \right ] \right \}^{N-2}
\end{equation}

Note that since the base distribution is different in the stochastic interpolant vs. the thermodynamic integration, the free energy differences are related by $\Delta F_{\text{TI}} = \Delta F_{\text{SI}} \,\,-\ln Z_{\text{ang}}$.

We sample from the base distribution by first drawing $N-1$ samples from the bond length distribution via rejection sampling with a Gaussian envelop prior centered at the maximum.
The cosine of the $N-2$ bond angles are then sampled from a truncated Gaussian distribution and the azimuth is taken as uniform over the range $[0,2\pi)$.
The first particle is placed at the origin and the chain is built sequentially.
The parameters $\kappa$ and $\gamma$ are chosen to simultaneously match the average squared end-to-end distance and the variance of the bond angles observed in the target samples.

\subsubsection{EGNN Node and Edge Features}

We again use the EGNN model of \cite{Satorras2021} to parameterize the velocity and score fields, though we augment it in several ways compared to the Lennard-Jones application above.
First, we expand the node features to include a half-period Fourier embedding of both the time and the normalized index of the particle:
\begin{equation}
    \textbf{h}_i = \left [ t, (\cos k \pi t),(\sin k\pi t)\right ] \oplus \left [ s/N,(\cos j \pi s/N),(\sin j\pi s/N) \right ]
\end{equation}
where $t$ is the time, $s=i-1$ is the (shifted) index/position of the particle along the chain (starting at zero), and the trigonometric terms include harmonics $1 \leq k \leq n_t$ and $1 \leq j \leq n_s$, so that $\textbf{h}_i$ has dimension $d_{\text{h}}=2+2n_t+2n_s$.

Second, we introduce a new edge feature to describe bonded interactions.
The edge features in the original network are simply the squared distances between particle positions, which are already passed into the message-generating network and are therefore redundant.
We replaced these redundant features with a simple flag for each edge, taking a value of $1$ if the particles are connected by a bonded potential and zero otherwise, i.e. a Kronecker delta:
\begin{equation}
    \mathbf{e}_{ij} = \delta_{|i-j|,1}
\end{equation}
By encoding this information directly, we ease the demands on the model, which would otherwise have to learn the distinction between bonded and nonbonded particles independently.
The network is otherwise identical to that of the LJ system.

\subsubsection{Training and Hyperparameters}

The equivariant optimal transport method outlined above was used with a minibatch size of 256.
Training was carried out using the Adam optimizer with a learning rate of 1e-3 for 200,000 iterations followed by a series of (100k,25k,25k,25k,25k) iterations at rates of (1e-4,3e-5,1e-5,3e-6,1e-6), respectively.
The same exponential moving average (EMA) scheme applied in the case of the LJ system was also utilized here.

\subsection{Computing Infrastructure}\label{sec:infra}

All model training and inference was conducted on a single NVIDIA RTX PRO 2000 Blackwell.
Monte Carlo and Brownian dynamics simulations (along with associated thermodynamic integrations) were conducted on an Intel Core Ultra 7 265 processor distributed across 16 cores.

\bibliography{refs}

\end{document}